\documentclass[twocolumn,amsmath,amssymb,aps,showkeys]{revtex4-2}
\usepackage{float}
\usepackage{caption}
\usepackage{subcaption}
\usepackage{graphicx}
\usepackage{dcolumn}
\usepackage{bm}
\usepackage{booktabs}

\begin{document}
\title{Prediction of Magnetic State of UO$_2$ within Hubbard-corrected Density-Functional Theory: A self-consistent approach}
\author{Mahmoud Payami}
\email{mpayami@aeoi.org.ir}

\affiliation{School of Physics \& Accelerators, Nuclear Science and Technology Research Institute, \\ AEOI, 
	P.~O.~Box~14395-836, Tehran, Iran}

\begin{abstract}
The magnetic state of UO$_2$ was determined experimentally to be anti-ferromagnetic. Starting from this experimental fact, researchers have calculated other properties within the Hubbard-corrected density-functional theory, DFT+U. Up to now, the Hubbard parameters for UO$_2$ were usually so chosen that the calculations give good results for some experimental data. 
Also, to our knowledge there exists no valid theoretical research report on the energetically stable magnetic state of this system. In present work, employing the new method which is based on density-functional perturbation theory, we have determined self-consistently the Hubbard parameters and ground-state energies for UO$_2$ crystal in both ferromagnetic and anti-ferromagnetic configurations, and the calculated results show that UO$_2$ crystal energetically favors an anti-ferromagnetic state with a small energy difference.
In all the calculations the PBE-sol approximation was used for the exchange-correlation energy functional.    
\end{abstract}


\maketitle

\section{Introduction}\label{sec1}

The popular local-density approximation (LDA) \cite{kohn1965self,perdewzunger81} and generalized gradient approximation (GGA) \cite{gga-pbe1996} approximations for the exchange-correlation (XC) energy functional in the density-functional theory (DFT) \cite{hohenberg1964,kohn1965self} suffer from self-interaction errors, which are significant in systems containing atoms with localized {\it d} and {\it f} orbitals, and by over-delocalization of their corresponding wave-functions lead to incorrect prediction of metallic behavior for Mott insulators. A simple and low-cost workaround is using the Hubbard-corrected DFT model to correct the correlation energy of localized orbitals in DFT energy-functional, DFT+U, in which only on-site corrections are added to the DFT energy functional \cite{coco-degironc2005}: 

\begin{equation}\label{eq1}
E_{\rm DFT+U}=E_{\rm DFT}[n({\bf r})] + E_{\rm Hub}[{n_m^{I\sigma}}]-E_{\rm dc}[n^{I\sigma}],
\end{equation}
where $n({\bf r})$ is electron density, $n_m^{I\sigma}$ are orbital occupation numbers of atom at lattice site ${\bf R}_I$, and $n^{I\sigma}=\sum_m n_m^{I\sigma}$. The last term in right hand side of Eq.~(\ref{eq1}) is needed to avoid double counting of interactions contained in the first and second terms.
The rotationally invariant form \cite{dudarev1998} of the correction is given by \cite{coco-degironc2005}: 

\begin{equation}\label{eq2}
E_{\rm U}[{n_{m m^\prime}^{I\sigma}}]\equiv E_{\rm Hub}-E_{\rm dc} =\sum_{I,\sigma}\frac{U^I}{2} {\rm Tr[{\bf n}^{\it I\sigma}({\bf 1} - {\bf n}^{\it I\sigma}})],
\end{equation}
in which ${\bf n}^{I\sigma}$ is the atomic occupation matrix. This on-site correction, significantly improves the over-delocalization and lead to correct insulating properties.
The coefficients $U^I$ are called Hubbard on-site parameters, and for a known material, these $U^I$ values may be so adjusted that the calculations results well agrees with some experimental data \cite{payami-lattice-eg}. 
However, in the case of designing new materials there is no experimental data to be used for parameters fitting and on the other hand, it is very important for the theory describing a material to be a parameter-free one. The first attempts in this way was using linear-response constrained-DFT (LR-cDFT) within super-cell method \cite{coco-degironc2005,campojr2010} which was somewhat inconvenient and computationally demanding. Another scheme to estimate the parameters was named as constrained random-phase approximation (cRPA) \cite{Aryasetiawan2006} which was recently used by others \cite{Dudarev2019} to estimate the U parameter for uranium dioxide. 
However, a new method based on density-functional perturbation theory (DFPT) was recently introduced   which instead of using previous super-cells, focuses on the unit-cell, which is more convenient and relatively fast \cite{tim2018,tim2021,tim2022}. Using this new method, we have determined self-consistently the Hubbard parameters for UO$_2$ crystal in both ferromagnetic (FM) and ant-ferromagnetic (AFM) configurations of uranium atoms and showed that energetically UO$_2$ crystal favors an AFM magnetic state with a small energy difference, which is in agreement with experimental findings.
In our recent work \cite{payamisheykhibasaadat2023} we have demonstrated that choosing PBE-sol approximation \cite{gga-pbesol2008} for the XC functional leads to excellent results from self-consistent DFT+U calculations, and therefore we have employed it in all computations throughout this work.

The organization of this paper is as follows: Section~\ref{sec2} is dedicated to the computational details; in Section~\ref{sec3} we present and discuss the calculated results; and finally section~\ref{sec4} summarizes and concludes this research.

\section{Computational details}\label{sec2}
\begin{figure}
	\centering
	\includegraphics[width=0.5\textwidth]{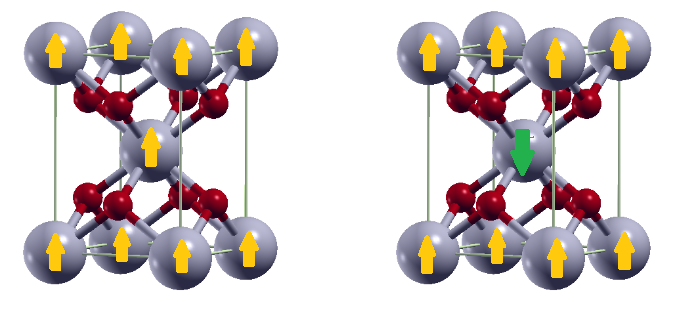}
	\caption{UO$_2$ crystal structure as simple tetragonal with six atoms basis. Left and right figures schematically represent FM and AFM configurations, respectively. Large grey and small red balls represent uranium and oxygen atoms, respectively. The up-spin and down-spin atoms are shown with yellow and green colors, respectively.}
	\label{fig1}
\end{figure}

The crystal structure of uranium dioxide is described by a simple tetragonal lattice with a six-atoms basis, shown in Fig.~\ref{fig1}. In FM configuration, all spins of U atoms have the same direction along $z$-axis, while to setup AFM structure for U atoms, we use the simple model in which the planes of U atoms alternate their spins along $z$ direction, i.e., we assume a 1-dimensional AFM.  

\begin{figure}
	\centering
		\includegraphics[width=0.5\textwidth]{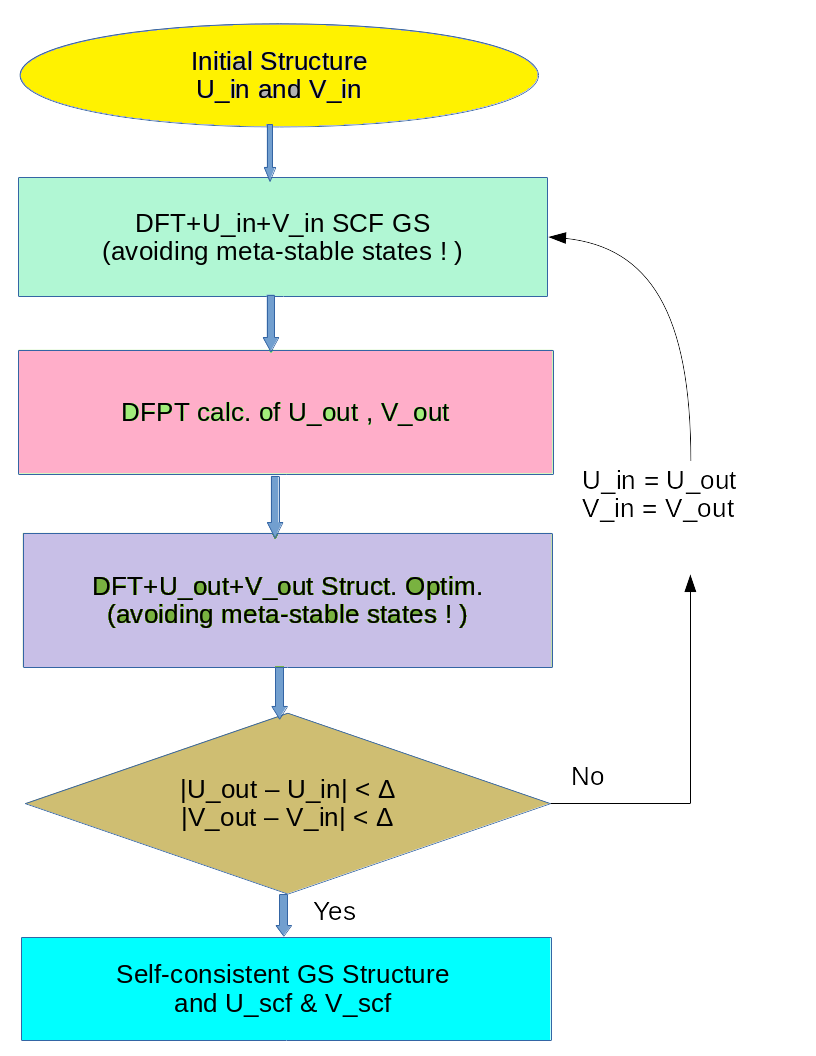}
	\caption{Flowchart of SCF determination of Hubbard parameters. In the first-step SCF and last-step structure-optimization, the meta-stable states were avoided \cite{payami-spinbroken2021,payami-smcomc2023}}
	\label{fig2}
\end{figure}

The Hubbard on-site U parameters were calculated self-consistently using the HP code \cite{tim2022} included in the Quantum-ESPRESSO code package \cite{qe-2009,qe-2020} for both FM and AFM configurations of U atoms (shown in Fig.~\ref{fig1}) in the context of PBE-sol \cite{gga-pbesol2008} approximation to the XC. The calculations include results for both "atomic" and "ortho-atomic" types of projections onto Hubbard orbitals.
The electronic structure calculations are based on the solution of the KS equations using the Quantum-ESPRESSO code package. For U and O atoms the scalar-relativistic ultra-soft pseudo-potentials (USPP) were used which were generated by the {\it atomic} code and generation inputs from the {\it pslibrary} \cite{DALCORSO2014337}, at https://github.com/dalcorso/pslibrary. The valence configurations U($6s^2,\, 6p^6,\, 7s^2,\, 7p^0,\, 6d^1,\, 5f^3 $) and O($2s^2,\, 2p^4 $) were adopted in the USPP generation.
Kinetic energy cutoffs for the plane-wave expansions
were chosen as 90 and 720~Ry for the wave-functions and densities, respectively. The smearing method of Marzari-Vanderbilt \cite{mv-smear1999} for the occupations with a width of 0.01~Ry were used. 
For the Brillouin-zone integrations in geometry optimizations, a $8\times 8\times 6$ grid were used;  All geometries were fully optimized for total residual pressures on unit cells to within 0.5 kbar, and residual forces on atoms to within 10$^{-3}$~mRy/a.u.
To self-consistent determination of the Hubbard parameters we have employed the HP code \cite{tim2022} following the flowchart shown in Fig.~\ref{fig2}. 

To start the self-consistent procedure for determining the Hubbard parameters according to Fig.~\ref{fig2}, the convergent q-mesh of $4\times 4\times 3$ for linear response calculations was adopted and we gave initial values for $U_{in}$ in each of FM and AFM configurations; for the initial structure we chose simple tetragonal structure with appropriate lattice constants consistent with cubic structure of side 5.47$\AA$. To avoid meta-stable states, we determined appropriate occupations of Hubbard orbitals $5f$ of uranium atoms \cite{payami-spinbroken2021,payami-smcomc2023}.  
In this second step, we start the DFPT calculation and obtain new values for parameters named as $U_{out}$. In the third step, using the parameter $U_{out}$ we obtained in the second step, the geometry of the system is optimized taking care of meta-stable states. In each cycle the differences between input and output parameters were monitored to see if the self-consistency is reached within $\Delta$ value. For this system the self-consistency was reached within 6 to 8 cycles in the flowchart with $\Delta < 10^{-4} $.  

\section{Results and discussions}\label{sec3}
The calculations were performed at the level of DFT with on-site Hubbard corrections (DFT+U) for both FM and AFM configurations, and the results showed that both FM and AFM magnetic states are insulators with different values for electronic band gaps. The self-consistent Hubbard parameters and geometric equilibrium lattice constants for FM and AFM states differ slightly; but their respective band gaps show relatively significant differences. The results are summarized in Table~\ref{tab1}.

\begin{table*} 
	\centering
	\caption{Self-consistent Hubbard parameters $U_{sc}$ in eV; total energy with respect to the lowest one corresponding to ortho-atomic AFM, per formula unit $\Delta E$ in eV; total and absolute magnetizations per formula unit in Bohr-magneton; equilibrium lattice constants in $\AA$; and electronic band-gaps in eV for different cases studied. PBE-sol approximation was used for the XC, which turns out to be the best one.}
	\begin{tabular}{@{}lccccccccccccc@{}}\toprule
		& \multicolumn{6}{c}{AFM} &\phantom{abc} & \multicolumn{6}{c}{$\;\;\;\;\;\;\;$FM} \\
		\cmidrule{2-7} \cmidrule{9-14}
		U-proj& $U_{sc}\;(eV)$ & $\Delta E$ & $M_{tot}$& $M_{abs}$ & $a\;(c)\;(\AA)$ & $E_g\;(eV)$ && $U_{sc}\;(eV)$ & $\Delta E$ & $M_{tot}$& $M_{abs}$ & $a\;(c)\;(\AA)$ & $E_g\;(eV)$\\ \midrule
		o-atomic & 2.9787 & 0.0000 & 0.00 & 2.150 & 5.4540(5.4631) & 2.2281 && 2.9936 & 0.0103 & 2.00 & 2.175 &
		5.4541(5.4655) & 1.7102 \\
		atomic   & 2.0862 & 0.0961 & 0.00 & 2.155 & 5.4546(5.4766) & 1.3699 && 2.1247 & 0.1140 & 2.00 & 2.190 &
		5.4582(5.4770) & 0.9162 \\ \bottomrule 
	\end{tabular}\label{tab1}
\end{table*}

As is seen from Table~\ref{tab1}, the AFM configuration with orthogonalized projections on Hubbard orbitals is the lowest energy state, and therefore the energetically stable state of UO$_2$ is an AFM configuration. To our knowledge, this result is obtained for the first time. The lattice constants differ by $0.02 \AA$  and all are in excellent agreement with experiment. The band gaps for AFM are larger than corresponding values in FM configurations. It is important to note that the self-consistent Hubbard parameters are different for AFM and FM configurations which result from different responses of them to an external perturbation. Therefore, applying the same empirical value for AFM and FM states of a system and comparing the energies to decide the favored state is shown to be incorrect. 
As is seen, the band gap of AFM with orthogonalized projections gives the best agreement with experiment \cite{schoenes1978optical}.

\begin{figure*}
	\centering
	\includegraphics[width=0.8\textwidth]{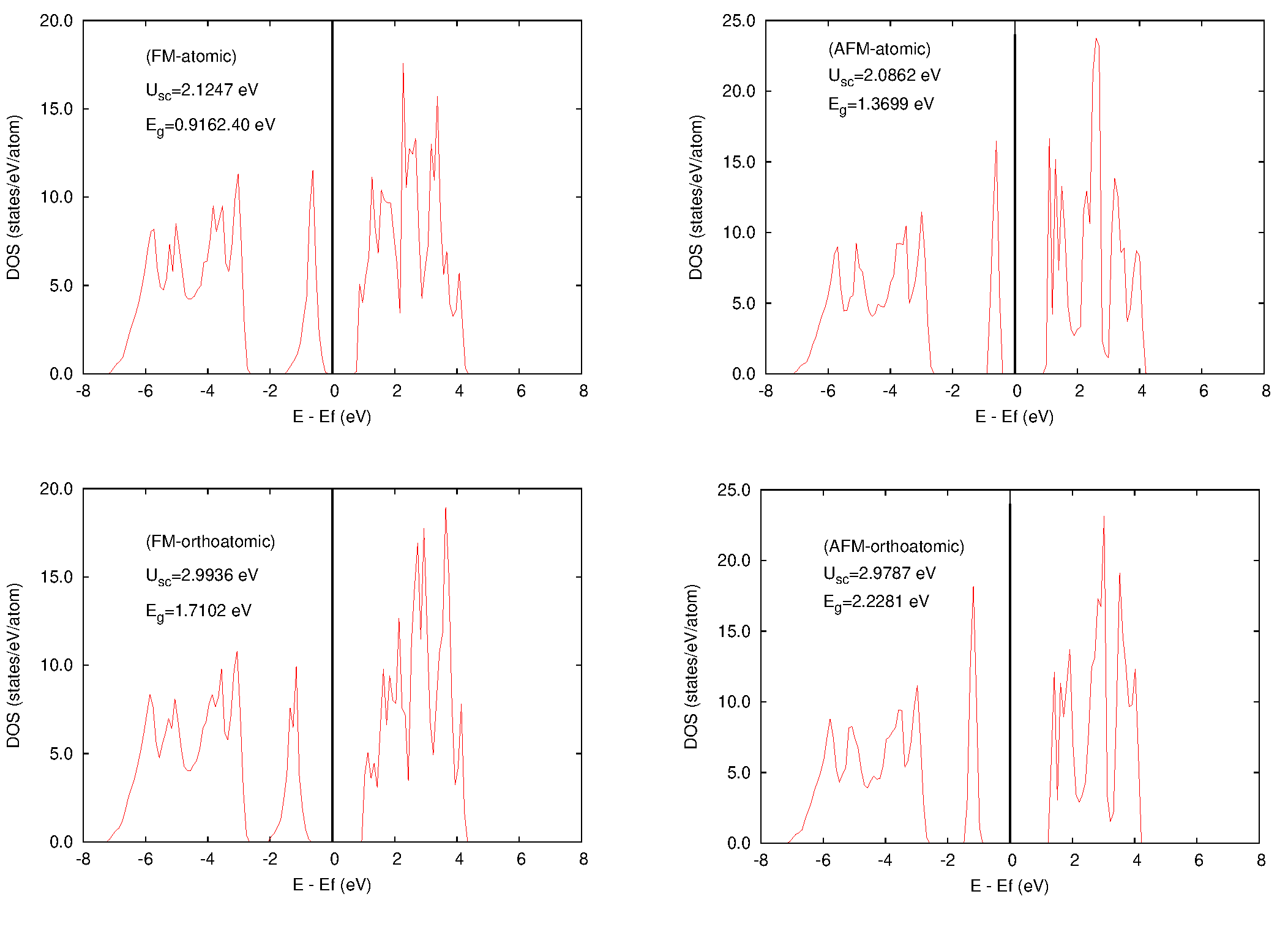}
	\caption{Total electronic density of states (DOS) for AFM and FM configurations of UO$_2$ crystal. It is observed that in a given magnetic state of UO$_2$, the band gap is larger for orthogonalized projections. Moreover, the band gaps of AFM are larger than corresponding values for FM.}
	\label{fig3}
\end{figure*}

To compare the overall electronic states of FM and AFM configurations, we have plotted the electronic total density of states (DOS) and shown in Fig.~\ref{fig3}. Inspecting the Fig.~\ref{fig3}, one notices that the behaviors near the band edges are different for two magnetic states, and the value of band gap for FM is smaller than that of AFM.


\section{Conclusions}\label{sec4}
In the study of strongly-correlated UO$_2$ system, usually one takes an AFM configuration for the uranium atoms which is borrowed from experimental findings, and the best job afterwards is to calculate the Hubbard parameter U self-consistently, as done by present authors in earlier work. However, it is very much interesting to predict the magnetic state theoretically as well without resorting to experimental facts. This becomes vital in designing new novel materials. 
In present work, using DFPT, we have calculated self-consistently the Hubbard on-site parameters and thereof the total energies for the two AFM and FM states of UO$_2$ crystal and have shown that the resulting parameters are different and AFM configuration is energetically favored by a small value of $0.01 eV$/(formula unit).  


\section*{Acknowledgement}
This work is part of research program in School of Physics and Accelerators, NSTRI, AEOI.  

\section*{Data availability }
The raw or processed data required to reproduce these results can be shared with anybody interested upon 
sending an email to M. Payami.

\end{document}